\documentclass[referee,a4]{raa}
\usepackage{graphicx,times}
\usepackage{natbib}
\usepackage{amssymb,amsmath}
\bibpunct{(}{)}{;}{a}{}{,}

\usepackage[a4paper=true,dvipdfm=true,pagebackref=true]{hyperref}
\hypersetup{pdftitle = The title of my PDF, pdfauthor = My name, pdfsubject= The subject, pdfkeywords = keyword1 keyword2 keyword3}
\hypersetup{colorlinks = true, linkcolor = green, anchorcolor = red, citecolor = blue, filecolor = red, pagecolor = red, urlcolor = red}

\newcommand{\Msolar}{\mbox{\,$\rm M_{\odot}$}} 
\newcommand{\Rsolar}{\mbox{\,$\rm R_{\odot}$}} 
\newcommand{\Lsolar}{\mbox{\,$\rm L_{\odot}$}} 
\def\simge{\mathrel{\raise1.16pt\hbox{$>$}\kern-7.0pt \lower3.06pt\hbox{{$\scriptstyle\sim$}}}} 
\def\simle{\mathrel{\raise1.16pt\hbox{$<$}\kern-7.0pt \lower3.06pt\hbox{{$\scriptstyle \sim$}}}} 

\begin{document}

\title{The double helium-white dwarf channel for the formation of AM\,CVn binaries}

 \volnopage{ {\bf 2012} Vol.\ {\bf X} No. {\bf XX}, 000--000}
   \setcounter{page}{1}

   \author{Xianfei Zhang\inst{1}, Jinzhong Liu\inst{2}, C.~Simon Jeffery
      \inst{3,4}, Philip D.~Hall\inst{3}, Shaolan Bi\inst{1}
   }

   \institute{ Department of Astronomy, Beijing Normal University, Beijing, 100875, China; {\it zxf@bnu.edu.cn}\\
\and
     Xinjiang Astronomical Observatory, Chinese Academy of Sciences, 150 Science 1-Street, Urumqi, Xinjiang 830011, China\\
\and
Armagh Observatory and Planetarium, College Hill, Armagh BT61 9DG, UK\\
\and
School of Physics, Trinity College Dublin, Dublin 2, Ireland\\
\vs \no
   {\small Received XXXX; accepted XXXX}
}

\abstract{Most close double helium white dwarfs will merge within a Hubble time due to orbital decay by gravitational-wave radiation.
However, a significant fraction with low mass ratios will survive for a long time as a consequence of stable mass transfer.
Such stable mass transfer between two helium white dwarfs (HeWD) provides one channel for the production of AM\,CVn binary stars.
In previous calculations of double HeWD progenitors, the accreting HeWD was treated as a point mass.
We have computed the evolution of 16 double HeWD models in order to investigate  the consequences of treating the detailed evolution of both components.
We find that the boundary between binaries having stable and unstable mass transfer is slightly modified by this approach.
By comparing with observed periods and mass ratios, we redetermine masses of eight known AM\,CVn stars by our double
HeWDs channel, i.e. HM\,Cnc, AM CVn, V406\,Hya, J0926, J1240, GP\,Com, Gaia14aae and V396\,Hya.
We propose that central spikes in the triple-peaked emission spectra
of J1240, GP\,Com and V396\,Hya and the surface abundance ratios of N/C/O in GP\,Com
can be explained by the stable double HeWD channel.
The mass estimates derived from our calculations are used to discuss
the  predicted gravitational wave signal in the context of the Laser Interferometer Space Antenna (LISA).}


\authorrunning{X. Zhang et al. }            
\titlerunning{2$\times$HeWD evolution and AM\,CVn stars}  

\maketitle

%
\section{Introduction}           

Most stars are members of binary star systems. White dwarfs represent the end
state of more than $90\,\%$ of all stars.
After the components of close binaries expand, interact, and accrete and/or lose mass,
most binaries end their lives containing at least one white dwarf (WD).
Double white dwarfs are particularly interesting.
They are an important source of gravitational wave emission,
which results in orbital decay. The subsequent merger of the
components can produce a type Ia supernova when the product is sufficiently massive, i.e. the total mass is close to or greater than Chandrasekhar mass.
Close double white dwarfs are formed following one or more episodes in which
a common envelope forms around both stars, and is ejected from the binary system
\citep{Webbink84,Iben97,Han98,Saio00,Nelemans01}.

Among the double white dwarfs, $53\,\%$ are double helium white dwarfs (2$\times$HeWDs), which have component masses less than $0.5\Msolar$ \citep{Nelemans00}.
Most  helium white dwarfs have masses in the range $0.25\Msolar$ to $0.5\Msolar$.
However, extremely low-mass HeWDs ($<0.25\Msolar$) have been found in
increasing numbers in recent years.
Since these cannot come from single-star evolution within a Hubble time, it is
assumed they are a product of  binary star evolution \citep{Kilic10,Kilic11,Brown2016,Chen2017}.

Orbital decay by gravitational-wave radiation
will cause most close 2$\times$HeWDs to merge within a Hubble time.
Most of the merged products will evolve to become hot subdwarfs
located close to the helium main sequence in the Hertzsprung--Russell (HR) diagram \citep{Iben90,Saio00,Zhang12a}.
Not all 2$\times$HeWDs will merge.

There are two key requirements for a close binary white dwarf to merge.
The first one is that gravitational radiation makes the binary lose
orbital angular momentum so that the stars move closer to each other.
The rate of loss of orbital angular momentum \citep{Landau62} is expressed as

\begin{equation}
\frac{\dot{J}_{\rm orb}}{J_{\rm orb}} =
-8.3\times 10^{-10} \times \left( \frac{M_1}{\Msolar}\right) \left(\frac{M_2}{\Msolar}\right)
\left(\frac{M_1+M_2}{\Msolar}\right)\left(\frac{a}{\Rsolar}\right)^{-4}{\rm yr}^{-1},
\end{equation}
where $M_1$ and $M_2$ are the masses of stars in the binary.

The second is that the mass ratio $q=M_2/M_1$  should be greater than some
critical value $q_{\rm crit}$, where $M_1>M_2$ are the masses of the white dwarfs, being accretor and donor respectively.
If, at the point where the larger (less massive)
white dwarf fills its Roche  lobe and starts to lose mass,
   \begin{equation}
    q \equiv \frac{M_2}{M_1} \geq q_{\rm crit}  \equiv \frac{5}{6}+ \frac{\zeta(M_2)}{2},
    \end{equation}
where $\zeta(M_2) \equiv {\rm d} \ln R_2/ {\rm d} \ln M_2$ is obtained from the white-dwarf mass--radius
relation, its radius will increase more quickly than the
separation will increase due to the transfer of angular momentum.
This leads to unstable (runaway) mass
transfer on a dynamical timescale.
By assuming a simple mass-radius relation,
i.e. $R_2 \propto M_2^{-1/3}$, a value $q_{\rm crit}=2/3$
is obtained.
This result has also been demonstrated from numerical simulations \citep{Motl07}
i.e. white dwarf binaries with mass ratios
$2/3 < q <  1$ will merge dynamically;
stable mass transfer will occur for $q \leq 2/3$, possibly leading to the formation of an AM\,CVn star.

AM\,CVn stars are interacting double stars with white dwarf accretors,
with optical spectra dominated by helium, and orbital periods less than about one hour.
The white dwarf accretes helium-rich material from a low-mass donor  \citep{Solheim2010}.
They are faint, blue and variable, and most are identified by their helium-dominated emission-line spectra.
Two formation channels have been proposed:
(1) stable mass transfer between two white dwarfs;
(2) stable mass transfer from helium stars to white dwarfs.
The detailed population synthesis of AM\,CVn stars was calculated by \citet{Nelemans2001}.
As compact short-period binaries, their potential gravitational-wave emission has subsequently computed by several groups including \citet{Nelemans2004,Ruiter2010,Yu2010} and most recently by \citet{Kremer2017}.

The response of compact accretors to stable mass transfer has not been included in detail in previous calculations of AM\,CVn binaries.
This omission may have non-negligible consequences.

In this paper, we compute detailed 2$\times$HeWD models by evolving both stars
during the stable mass transfer phase.
We aim to identify the conditions that lead to the formation of AM\,CVn stars,
and to compare the results with observed stars that may have evolved through this channel.
As  AM\,CVn systems represent an important source of observable gravitational waves (GW) we also investigate how our new calculations affect the GW predictions.

\begin{figure}
\centering
\includegraphics [angle=0,scale=0.8]{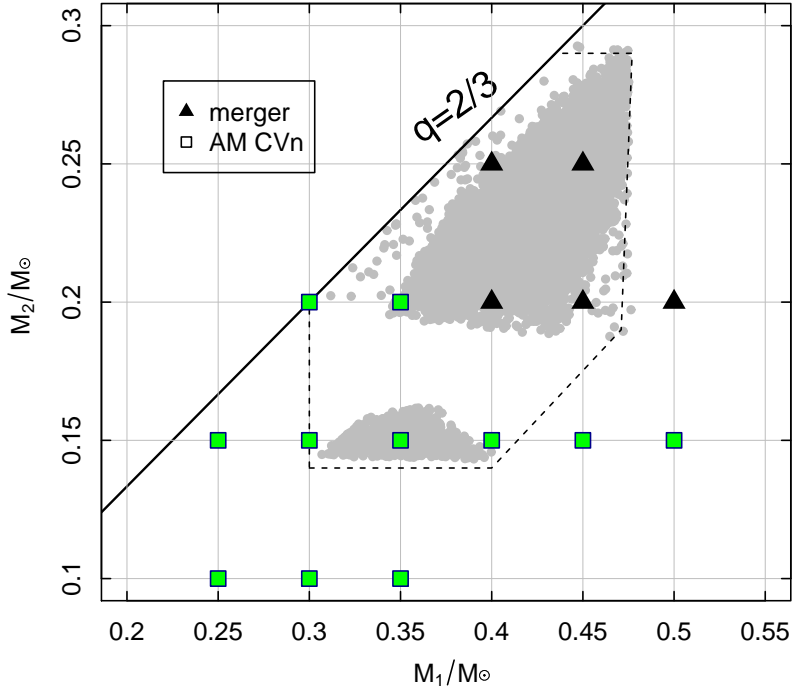}
\caption[]{Models of double helium white dwarfs.
    The symbols indicate which models form AM\,CVn stars (squares), and which form mergers (triangles).
    The grey dots indicate the locations of 2$\times$HeWD models with $q<2/3$ from a simulation of 10 million binary stars   which  either merge or reach a semidetached state within $13\,\rm{Gyr}$.}
\label{modelfig1}
\end{figure}

\section{Models}
To investigate the stable mass transformation of two helium white dwarfs,
16 pairs of HeWD models have been calculated on a grid with increments of $0.05\Msolar$ in mass,
as shown in Fig.~\ref{modelfig1}.

In our calculations, we treat the evolution of both white dwarfs in detail, using the binary module of the stellar evolution code
\texttt{MESA} \citep[Modules for Experiments in Stellar Astrophysics v8118;][]{paxton11,paxton13,paxton15}.
This avoids the approximation of treating the accretor as a point mass.

To construct a helium white dwarf, starting with a $1.5\Msolar$ zero-age main-sequence star with metallicity of $Z=0.02$,
evolution is computed until the He core reaches the required mass, e.g. $0.30\Msolar$.
Nucleosynthesis is switched off and a high mass-loss rate is applied to
remove the hydrogen envelope. Low-mass helium white dwarfs
are considered to retain a small hydrogen envelope \citep{Althaus2001,Steinfadt2010,Hall2013}.
Mass loss is therefore switched off when the hydrogen envelope is reduced to $10^{-4}\Msolar$.
The model evolves straight to the WD cooling track and does not ignite helium at this stage,
i.e. $\log (L/\Lsolar) = -2$. This procedure produces the initial HeWD models which are used
as input to the binary module;  the evolution of both stars is then computed simultaneously,
with mass transfer rates and orbital parameters computed self-consistently. The mass transfer
is computed assuming optically thick overflow  \citep{Kolb90}, as also  used for
AM CVn calculations by \citet{Deloye2006}. In the optically thick overflow model,
the mass transfer rate is sensitive to the difference between the stellar  and Roche
lobe radii,  depending particularly on the detailed structure of the donor's outer layers.
The pressure scale height $H_{\rm p}$ in the outer layer of the donor is much smaller than
it's radius $R$, i.e. $H_{\rm p} \ll R$. 

Other details of the \texttt{MESA} physics are as follows.
The ratio of mixing length to local pressure scale height is set to $\alpha = l/H_{\rm p} = 1.9179$,
as found by the solar calibration of \citet{paxton11}.
The opacity tables are from \citet{Iglesias1996} and \citet{Ferguson2005}.
Because the abundances of carbon and oxygen in the interior change after a He flash,
we use the OPAL Type 2 opacity tables.
The outer boundary condition is chosen to be an Eddington gray photosphere.
Orbital-decay by gravitational-wave radiation have been switched on.
We do not consider rotational, semiconvective, thermohaline mixing, or mass loss due to a stellar wind.

The birth rate of AM\,CVn systems from the double white dwarf path is very sensitive to the synchronization timescale  ($\tau_s$) of the white dwarfs spin with the orbital motion \citep{Nelemans01}.
With $\tau_s < 1000 \rm yr$, the birth rate can increase by  more than a factor of 2 \citep{Marsh2004}.
This is related to whether accretion is direct or mediated through a disk \citep{Gokhale2007}.
A strong coupling i.e. $\tau_s \rightarrow 0$, is used in our calculation in order to produce more AM\,CVn systems.
We assume that mass transfer is conservative, so that the total mass and orbital angular momentum (apart from that lost by gravitational wave radiation) remains constant during evolution.

As the binary orbit decays due to gravitational wave radiation, 2$\times$HeWD systems
reach a semidetached state (in which the lower-mass component fills its Roche lobe), with minimum period about 5 minutes.
After mass transfer begins, evolution may proceed in one of two ways.
For a high mass ratio system, the accretor will expand and fill its Roche lobe soon after accretion has begun.
The two components will form a common envelope and then merge.
For a low mass ratio system, the accretor will remain detached and
stable mass transfer through the $L_1$ point will continue;  such
systems are identified with AM\,CVn binaries.
From the  \texttt{MESA} calculations, we make the assumption that a binary will merge when both stars completely fill their Roche lobes; the calculation is halted at this point.
These are shown as solid triangles in Fig.~\ref{modelfig1}.
For the  models which do not merge, we compute the detailed evolution including
 stable Roche lobe overflow (RLOF: squares in Fig.~\ref{modelfig1}).
The calculation is stopped when the surface luminosity of the donor drops to $\log (L/\Lsolar) = -5$.

For example, the $0.40$+$0.25\Msolar$ model will merge and form a hot subdwarf \citep[cf.][]{Zhang12a}.
However, the $0.40$+$0.15\Msolar$ model will survive and continue to transfer mass.
The orbital period will increase from $5\,\rm{min}$ to $\sim 40\,\rm{min}$ over a timescale of $1\,\rm{Gyr}$;
increasing to $\sim 50\,\rm{min}$ after a further $2$--$3\,\rm{Gyr}$.
Although the orbital periods of semi-detached 2$\times$HeWD binaries are small,
the orbital expansion due to mass transfer is faster than contraction due to GW radiation.
Thus, the orbit becomes wider during a long-lived evolution which can survive for a few Gyr.
In summary, there are two different evolutionary directions during mass transfer in a 2$\times$HeWD,
i.e. to merge or to form an AM\,CVn binary.
Those two directions depend on the initial HeWD masses and are shown in Fig.~\ref{modelfig1}.

There are some differences with previous calculations.
A mass transfer rate $\dot{M} = 10^{-5}\Msolar\,\rm yr^{-1}$ was assumed by \citet{Nelemans01} as the boundary between stable and unstable mass transfer.
Binaries with $\dot{M} > 10^{-5}\Msolar\,\rm yr^{-1}$  would have unstable mass transfer and then merge.
For example \citet{Nelemans01} find that a $0.40$+$0.25\Msolar$ model does not merge but reaches stable RLOF because $\dot{M} < 10^{-5}\Msolar\,\rm yr^{-1}$.
However, from detailed calculations for the same model ($0.40$+$0.25\Msolar$) we find that, although
the  maximum mass transfer rate $\dot{M}_{\rm max} = 6.3\times10^{-6}\Msolar\,\rm yr^{-1}$,
helium-burning in the accretor can be ignited. In this model, the accretor radius increases and the
binary components merge.
This result emphasises the need for the evolution of both accretor and donor to be included in the calculation.

To investigate AM\,CVn binaries formed through the 2$\times$HeWD channel,
we need to know the mass distributions of both helium white dwarfs.
We therefore compute the properties of a population of $10^7$ primordial binary systems
using a Monte Carlo approach.
Using the rapid binary evolution code (\texttt{BSE}, \citealt{Hurley00,Hurley02}) and the technique of population synthesis, we identify those binaries
which evolve to become 2$\times$HeWDs with $q<2/3$.

The \texttt{BSE} input parameters are the same as those used for
modelling the Galactic rate of double WDs by \citet{Han98} and \citet{zhang2014}.
We considered a single population of $10^7$ coeval binaries, and only counted the pairs of 2$\times$HeWDs
which will merge or become semidetached. In the Monte Carlo
simulation, all stars are assumed to be members of binaries and have circular orbits. The masses of the
primaries are generated according to the formula of \citet{Eggleton89}, {\it i.e.} they follow the initial
mass function of \citet{Miller79}  and in the mass range $0.08$ to $100\Msolar$. The secondary mass, also with
a lower limit of $0.08\Msolar$, is subsequently obtained assuming a constant mass-ratio distribution.
The distribution of orbital separations, $p(a)$, is that of \citet{Han98}:
\begin{equation}
  p(a) =
  \begin{cases}
    0.070(a/a_0)^{1.2}  & \phantom{a_0 \le} a \leq a_0 \\
    0.070              & a_0 \le a \le a_1, \\
  \end{cases}
\end{equation}
where $a_0=10\,\Rsolar$, $a_1=5.75 \times 10^6\,\Rsolar = 0.13\,\rm{pc}$.
According to this, approximately $50$ per cent of stellar systems have orbital periods greater than $100\,\rm {yr}$ \citep{Han98} and are here considered to be single stars.

We find that these models occupy a region outlined in Fig.~\ref{modelfig1}.
These results give a probability distribution of all possible mergers or semidetached HeWDs.
The majority have initial masses in the range $0.35$+$0.2$--$0.47$+$0.28\Msolar$.
Another concentration is found with masses in the range $0.3$+$0.14$--$0.4$+$0.16\Msolar$
The majority of white dwarfs in those two regions are formed after two phases of
common envelope (CE) ejection. However, binaries in the less massive zone require critical
initial conditions. The two stars must be very close to each other, the more massive star evolves faster
and forms a HeWD after the first CE ejection (RGB+MS).
The second CE phase (HeWD+RGB) must occur at the base of RGB phase while the red giant
has a very low-mass helium core, which then forms a second low-mass HeWD following CE ejection.
Moreover, the first formed HeWD must be massive enough to eject the high mass envelope of
the RGB star.  Hence, with a narrow region of initial mass ratios and separations, the number of 2$\times$HeWDs in this subgroup is very small. 

\section{Results}

Eleven of the 16 original 2$\times$HeWD models reach a stable RLOF phase.
Of these, five are in the region where \texttt{BSE} population synthesis predicts 2$\times$HeWD to occur, namely the models with initial HeWD masses of  $0.30$+$0.15$, $0.30$+$0.20$, $0.35$+$0.15$, $0.35$+$0.20$
and $0.40$+$0.15\Msolar$.
For each theoretical system, we record the mass ratio, the donor and accretor masses, the orbital radial velocity, and the gravitational-wave strain amplitude as a function of  orbital period, which is a proxy for elapsed time after the components come into contact.
These will be compared with observed binary systems.

\begin{figure}
\centering
\includegraphics [angle=0,scale=0.8]{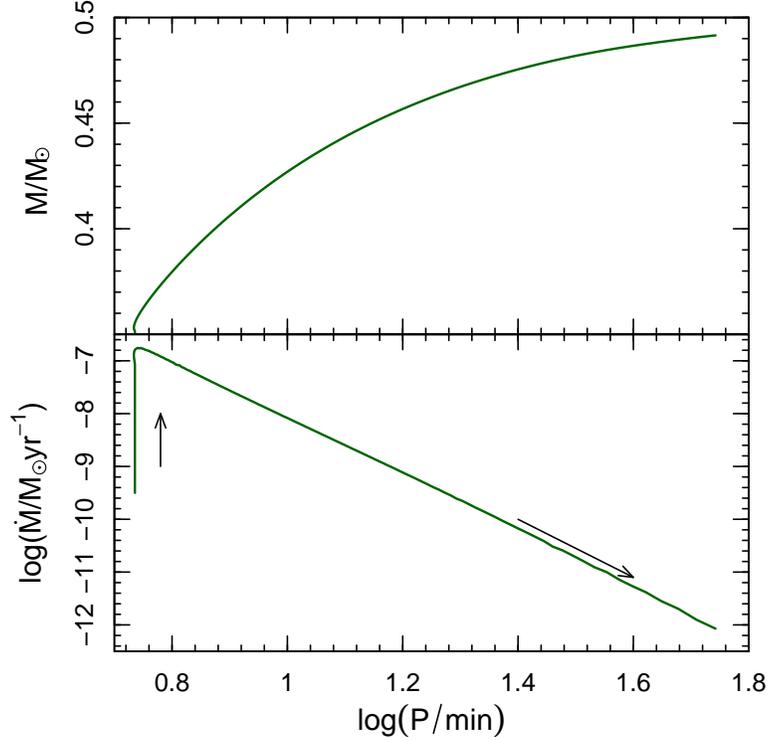}
\caption{The evolution of accretor's mass and mass transfer rate for model $0.35$+$0.15\Msolar$.
The arrows indicate the evolutionary direction.
 }
 \label{pevolfig2}
\end{figure}

Fig.~\ref{pevolfig2} shows the evolution of the $0.35$+$0.15\Msolar$ model.
The mass transfer rate  increases very sharply at the beginning of  Roche lobe overflow,
and then decreases exponentially as the orbital period increases. Meanwhile,
the accretor mass increases as the orbital period increases.

Mass is an important parameter for AM\,CVn systems.
However, AM\,CVn star masses are difficult to measure because the systems are
mostly  faint and the orbits are short, making the components difficult
to resolve. Mass determinations for most of AM\,CVn stars  depend on
assuming the formation channel, and has been estimated by comparing
observations with evolutionary models.
We are particularly interested in those AM\,CVn systems
which may have  formed through the 2$\times$HeWD channel.
In order to compare our models with observations, data for eight AM\,CVn stars
have been taken from \citet{Solheim2010,Campbell2015} and \citet{Kalomeni2016}.
These are: AM\,CVn itself, HM Cnc, J0926, V406\,Hya, J1240, GP Com, Gaia14aae and V396 Hya.
All have relatively precise mass ratios from either Doppler tomography or radial velocity
and eclipse measurements, but there are still no good measurements for their component masses.

Mass ratios have also been published for other AM\,CVn systems including CP Eri,  HP Lib, CR Boo, KL Dra, and V803 Cen.
In these cases, the mass ratios were estimated from an empirical superhump period -- mass ratio relation and have much larger systematic uncertainties.

\begin{figure}
\centering
\includegraphics [angle=0,scale=0.8]{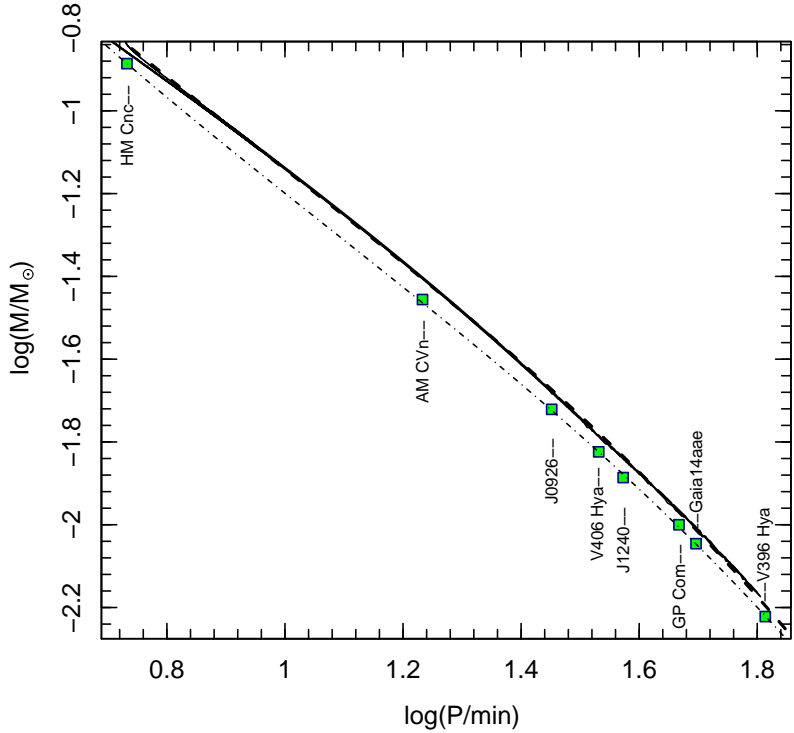}
\caption{The  evolution of five 2$\times$HeWD models showing the donor mass as a function of orbital period  (solid line).
The dashed line is the polynomial  fit  described in text.
The dot-dashed lines show the relation of \citet{Deloye2005}.
Minimum donor masses for eight selected AM\,CVn stars obtained using the method of \citet{Deloye2005}
are shown as squares.
 }
 \label{pmfig3}
\end{figure}

\begin{figure}
\centering
\includegraphics [angle=0,scale=0.8]{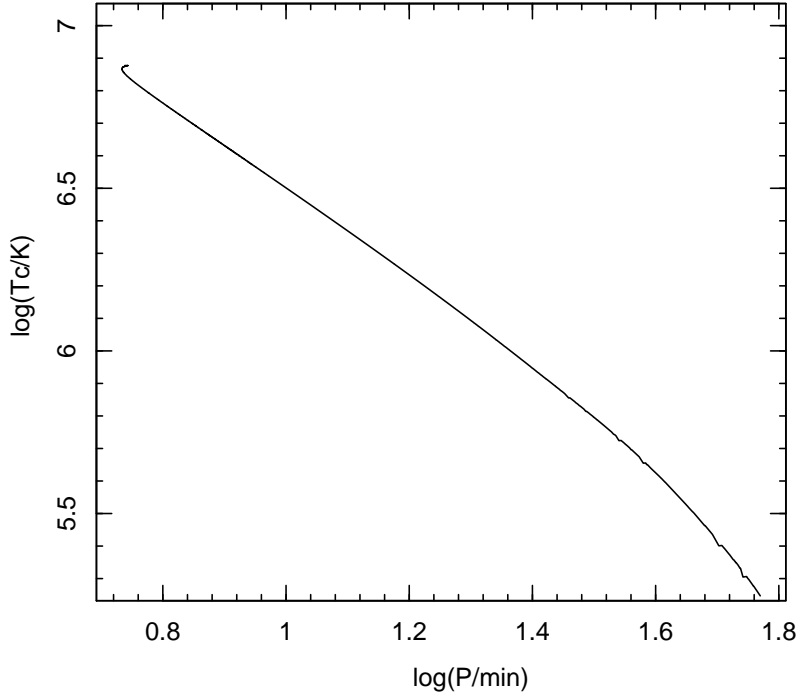}
\caption{The period-central temperature relation of model $0.35$+$0.15\Msolar$.}
 \label{ptcfig4}
\end{figure}

\begin{figure}
\centering
\includegraphics [angle=0,scale=0.8]{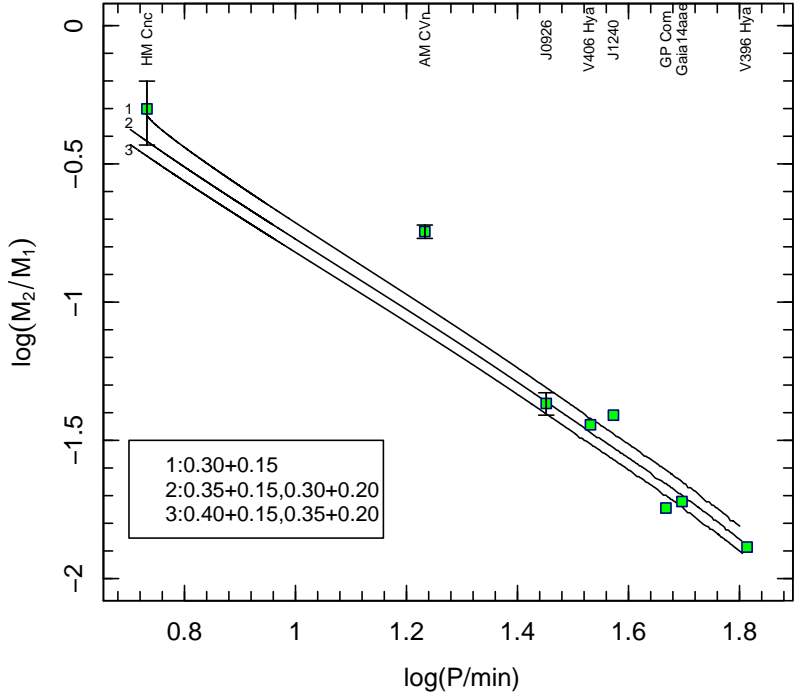}
\caption{The orbital-period--mass-ratio plane.
The evolution of five 2$\times$HeWD models showing the system mass ratio as a function of orbital period  (solid lines: component masses (\Msolar) shown in key).
The mass ratios of eight selected observed AM\,CVn stars, i.e. HM Cnc, AM\,CVn, J0926, V406 Hya, J1240, GP Com, V396 Hya \citep{Solheim2010} and Gaia14aae \citep{Kalomeni2016}, are shown by squares.}
 \label{pratiofig5}
\end{figure}

\subsection{Masses}

For stable mass transfer in AM\,CVn systems, the donor's stellar radius $R_2$ is assumed equal to its Roche lobe radius, $R_{\rm L}$, which can be approximated as \citep{Paczynski67}:
\begin{equation}
R_{\rm L} \approx 0.46 \,a \left(\frac{M_2}{M_1+M_2}\right)^{1/3}\,,
\end{equation}
when $q \equiv M_2/M_1 < 0.8$, and the orbit is circular.
Here $a$ is the orbital separation, $M_1$ and $M_2$ are the masses of the accretor and donor respectively.
From Kepler's Third law,
\begin{equation}
\frac{P^2}{a^3}= \frac{4 \pi^2}{G(M_1+M_2)},
\end{equation}
where $P$ is the orbital period.
Hence, equating $R_2=R_{\rm L}$,
\begin{equation}
P \propto R_2^{3/2} M_2^{-1/2}.
\end{equation}
Since white dwarfs' radii can be approximately related to their masses,
the donor mass $M_2$ is a function of orbital period.

Fig.~\ref{pmfig3} shows that direct measurements of $M_2$
provide  a strong constraint on the evolution models,
since the latter require that the donor mass $M_2$ must
be a function of orbital period. Helium white dwarfs must all follow
approximately the same mass--radius relation, regardless of age.
Thus, a single  $M_2$--$P$ relation correctly describes all 5 models computed,
as seen in the figure.

We obtain a polynomial fit to the theoretical $M_2$--$P$ relation for the
2$\times$HeWD channel:
\begin{equation}
    \rm log \left( \frac{M_2}{\Msolar} \right) =-1.7505 \times \rm log \left(\frac{P}{\rm min}\right)+0.6659 \times \rm log \left(\frac{P}{\rm min}\right)^2
    -0.2325\times \rm log\left(\frac{P}{\rm min}\right)^3+0.1758.
\end{equation}

For comparison, estimates of minimum donor masses have been taken from \citet{Deloye2005}.
The latter assumes a fully degenerate donor that fills its Roche lobe.
As \citet{Deloye2005} note, this assumption  may not always be satisfied since the thermal history (entropy) of the HeWD contributes strongly to the mass-radius relation used to derive the mass.
Nevertheless, these masses were also used by \citet{Bildsten2006} to calculate the heating of WDs by accretion, demonstrating that the latter significantly contributes to the optical and ultraviolet emission of AM\,CVn stars.
Fig.~\ref{pmfig3} demonstrates that the \citet{Deloye2005} minimum donor mass relation for AM\,CVn stars matches our 2$\times$HeWD relation well.
There remains a small difference between the relation
of \citet{Deloye2005} and our models. This difference is related to the mass-radius relations adopted.
The minimum donor masses from \citet{Deloye2005} are based on a mass-radius relation
of HeWDs with central
temperature $\log(T_{\rm c} / {\rm K})=4$. In our models, the central temperature is
higher than $\log(T_{\rm c} / {\rm K})=4$ and varies with time during evolution,
see the example shown in Fig.~\ref{ptcfig4}. The higher internal temperatures will lead to
larger radii and hence longer periods for a given mass.

It is therefore suggested that  our $M_2$--$P$ relation and the mass ratio $q$ could be
used to estimate both minimum donor and accretor masses with good accuracy.
Based on the $M_2$--$P$ relation, the donor masses can be obtained from the orbital periods.
Hence, the accretor masses may be calculated from the observed mass ratios.
By this means we obtain masses for HM Cnc ($0.289^{+0.101}_{-0.060}$+$0.144\Msolar$), AM CVn ($0.218^{+0.013}_{-0.011}$+$0.039\Msolar$),
J0926 ($0.493^{+0.051}_{-0.042}$+$0.021\Msolar$),
V406 Hya ($0.463$+$0.017\Msolar$), J1240 ($0.375$+$0.015\Msolar$), GP Com ($0.596$+$0.011\Msolar$),
Gaia14aae ($0.511$+$0.010\Msolar$) and V396 Hya ($0.490$+$0.006\Msolar$).
The major source of error is the observed mass ratio; only for HM Cnc, AM CVn and J0926 are errors given by \citet{Solheim2010}.

\begin{figure}
\centering
\includegraphics [angle=0,scale=0.8]{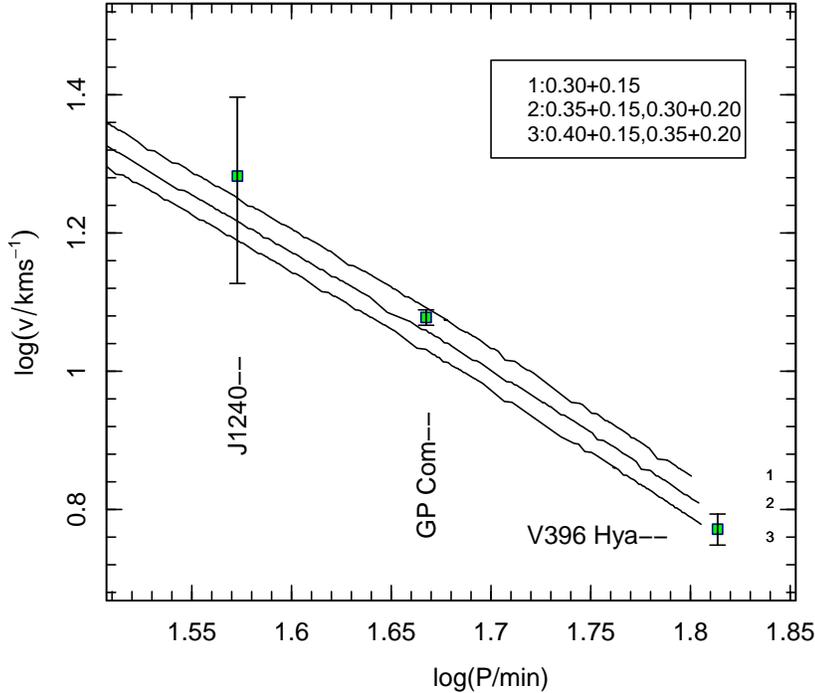}
\caption{The orbital-period--orbital-velocity plane.
The orbital velocity of the accretor is shown as a function of orbital period during the evolution of five 2$\times$HeWD models (solid lines: component masses (\Msolar) shown in key).
The squares with error bars indicate the velocity
of J1240 \citep{Roelofs2005}, GP Com and V396 Hya \citep{Kupfer2016}.}
 \label{vfig6}
\end{figure}

The donor's mass can written as a function of mass ratio and total mass of binary, i.e. $M_2=q/(1+q)\times M_{\rm total}$.
Hence, the mass ratio is a function of orbital period and total mass.
The evolution of mass ratio $q$ with orbital period $P$ for each of the five selected models is shown in  Fig.~\ref{pratiofig5}.
The model pairs of $0.30$+$0.20$ and $0.35$+$0.15\Msolar$, $0.35$+$0.20$
and $0.40$+$0.15\Msolar$ are share the same relations, because of the same total masses.
The mass ratios for seven of the eight AM\,CVn stars lie very close to the predicted $q$--$P$
relation for stable RLOF 2$\times$HeWD stars, i.e. HM Cnc, J0926, V406\,Hya, J1240, GP Com, Gaia14aae and V396\,Hya.
The AM CVn star was suggest to have a helium star donor \citep{Solheim2010}.

\subsection{Central spikes}
The stars J1240, GP Com and V396 Hya are classified as low-state AM\,CVn stars,
which show almost no outbursts, except for J1240 \citep{Levitan2015}.
They have very low mass transfer rates, in the range $10^{-13}$--$10^{-12}\,\Msolar\,\rm{yr}^{-1}$.
The principal difference between these three stars and other AM\,CVn stars is that their helium dominated spectra show lines with a triple-peaked profile.
This is interpreted as a double-peaked accretion disk profile plus a sharp,
low-velocity central component \citep{Marsh1999,Ruiz2001,Morales2003,Roelofs2005,Kupfer2016}.
The origin of the sharp central component, or ``central spike'', is unclear.
They  have only ever been observed in some AM\,CVn systems and
He-rich dwarf novae but never in hydrogen-dominated cataclysmic variables.
The central spikes have been suggested to originate from the white dwarf accretor or at least from near the accretor.
Moreover, the central spike shows significant radial velocity shifts as a function of orbital phase, which also indicate a likely origin close to the accretor, so that the semi-amplitude of the velocity shift should be equivalent to the orbital velocity of the accretor (assuming an orbital inclination close to $90^{\circ}$).

During mass transfer in a binary system; stars exchange
mass and angular momentum, which are here assumed to be conserved.
The total angular momentum is
\begin{equation}
    J =M_1 M_2 \left(\frac{Ga}{M_1+M_2}\right)^{1/2}.
\end{equation}
The change of separation due to mass transfer
\begin{equation}
   \frac{\delta a}{a} =2\frac{\delta M_2}{M_2} \left(\frac{M_2}{M_1}-1\right).
\end{equation}

In \texttt{MESA}, the orbital separation is calculated from the mass exchanged in each timestep of the
evolution.  Hence, the orbital periods are obtained from Kepler's laws;  the orbital velocities follow directly.

Fig.~\ref{vfig6} shows the orbital velocities $v$ of the accretor stars in our 2$\times$HeWD models as a function of $P$.
Observations generally provide the projected semi-amplitude, $K = v \sin i$, where $i$ is the inclination of the orbit to the plane of the sky. The minimum radial velocity of the central spikes of three
AM\,CVn systems are calculated with the estimated maximum inclination angle,
i.e. $53^\circ$, $78^\circ$, $79^\circ$ for J1240, GP Com and V396 Hya.
These agree well with the radial velocity of our accretor models.
However, to confirm central spikes originate from
the white dwarf accretor,  more accurate observations are still required.

\subsection{Abundance}
 \citet{Nelemans2010} argues that the surface abundances of AM\,CVn stars provide a good method to distinguish between AM\,CVn formation channels.
 However, it is very difficult to obtain such data.
 The only good observation is for GP\,Com where the abundance ratios  $\rm{N/C}>100$ \citep{Marsh1991} and $\rm{N/O}=8.8$ \citep{Strohmayer2004} by number.
From our calculations, all of the 2$\times$HeWD models show ratios $\rm{N/C}$ in the range $117$--$122$, and $\rm{N/O}$ in the range $8.4$--$8.6$, with almost no change in time.
Both are similar to the GP\,Com observation.
\citet{Bildsten2006} points out that the optical broadband colors and intensity of GP\,Com and V396\,Hya are as expected from a pure helium atmosphere white dwarf, providing supporting evidence for a 2$\times$HeWD origin for GP\,Com.

\begin{figure}
\centering
\includegraphics [angle=0,scale=0.8]{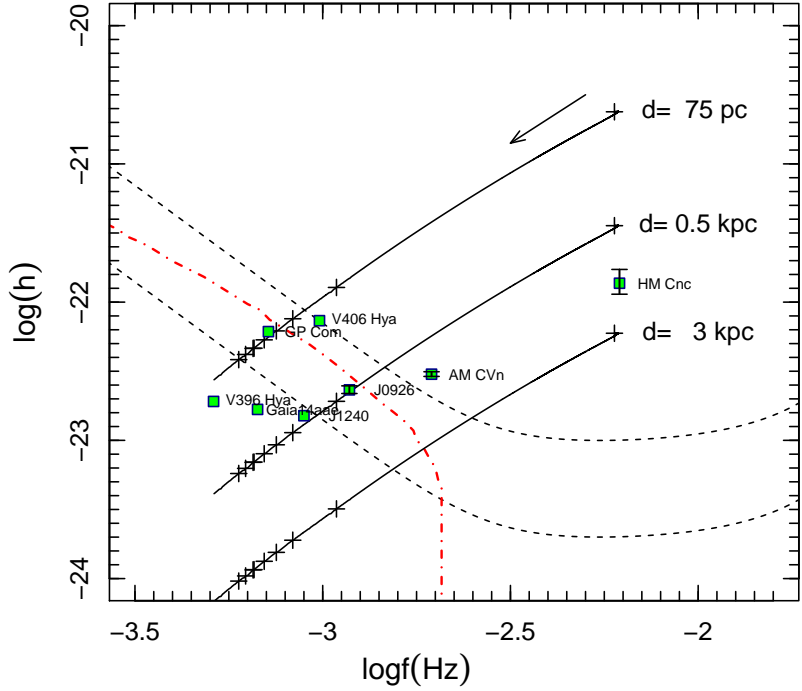}
\caption{The gravitational wave radiation frequency--strain amplitude plane.
The solid lines shows the results from our calculations of model $0.40$+$0.15\Msolar$.
The crosses denote intervals of $0.5\,\rm{Gyr}$ during stable RLOF evolution.
The squares shows the predicted positions of eight selected AM\,CVn stars.
The dashed lines show the LISA sensitivity for a one year mission giving a signal-to-noise ratio of $1$ and $5$ respectively.
The dot-dashed line indicate the average foreground
noise due to close white dwarf binaries \citep{Nelemans2004}.}
 \label{hfig7}
\end{figure}

\subsection{Gravitational waves}
AM\,CVn stars are an important source of gravitational wave (GW) emission and are expected to contribute strongly to the Galactic foreground detected by the Laser Interferometer Space Antenna (LISA) \citep{Nelemans2004}. The amplitude of a GW signal emitted by a binary system in a circular orbit  is given by
\begin{equation}
h = 0.5 \times 10^{-21}  \left( \frac{\mathcal{M}}{\Msolar}
\right)^{5/3} \! \! \! \left( \frac{P}{{\rm hr}} \right)^{-2/3}
\! \! \!\left( \frac{d}{{\rm kpc}} \right)^{-1},
\end{equation}
where $\mathcal{M} = (M_1 M_2)^{3/5} / (M_1 + M_2)^{1/5}$ is the chirp
mass, $P$ is the orbital period and $d$ is the distance to
the system \citep{Evans87,Yu2010}. The GW frequency $f = 2/P$.
To investigate the GW properties of our models, we take the $0.40$+$0.15\Msolar$ model as an example and calculate the GW strain amplitude for the model as it evolves from high frequency and high amplitude to low frequency and low amplitude (Fig.~\ref{hfig7}).
Three representative distance are considered, i.e. $75\,\rm{pc}$ (the distance of GP Com), $500\,\rm{pc}$ (most observed AM\,CVn stars are closer than this distance) and $3\,\rm{kpc}$ (a few more distant AM\,CVn stars).
The predicted negative shift in GW frequency (chirp) due to orbital expansion in accreting systems is well known \citep{Marsh2004,Kremer2017}.

We have recalculated the strain amplitude using the new masses of HM\,Cnc, AM CVn, J0926, V406\,Hya, J1240, GP\,Com, Gaia14aae and V396\,Hya predicted by our model (Section~3.1) and assuming lower limits on the distances of $1000$, $513$, $460$, $100$, $350$, $75$, $215$ and $92\,\rm{pc}$ respectively \citep{Solheim2010,Campbell2015}.
The strain amplitudes are shown in Fig.~\ref{hfig7}. With a one-year mission  and significance threshold of $1-\sigma$, up to five AM\,CVns could be detected by LISA.
Furthermore, more precise masses for the white dwarf components could be obtained from extended LISA measurements.
This is most likely for HM\,Cnc; the other AM\,CVn stars discussed here will be harder to resolve from the foreground of unresolved double white dwarf (DWD) binaries \citep{Nelemans2004,Yu2010,Liu2010}. For instance, the dot-dashed line shown in Fig.~\ref{hfig7}
indicates the average foreground noise due to unresolved close white dwarf binaries \citep{Nelemans2004}.
AM\,CVn stars close to or beneath such a line will be difficult to resolve from such a foreground.
Others, such as HM Cnc, AM CVn and V406 Hya are potentially detectable above the LISA noise threshold
{\it and} the DWD foreground.

\begin{figure}
\centering
\includegraphics [angle=0,scale=0.8]{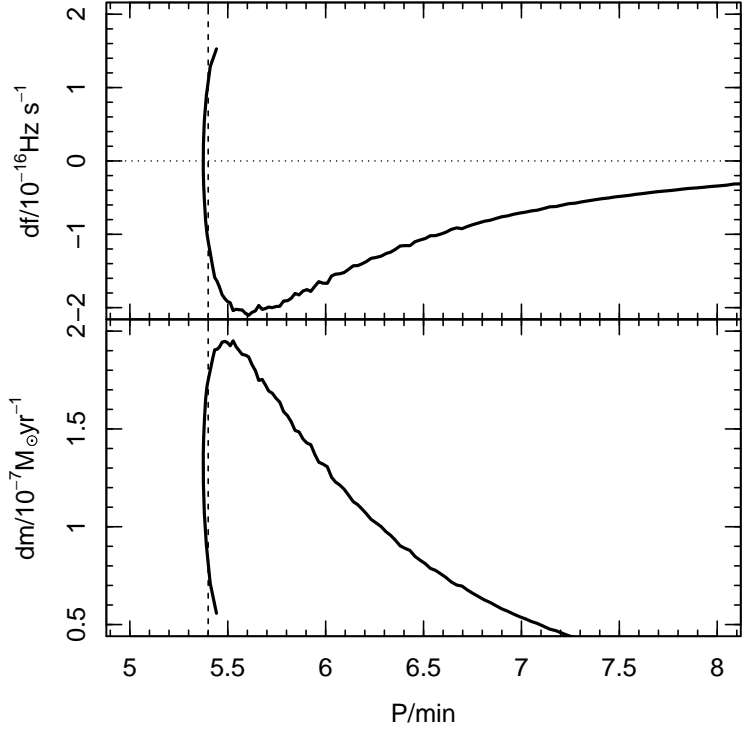}
\caption{The chirp and mass-transfer rate evolution with period for
model of $0.30$+$0.15\Msolar$. The period of HM\,Cnc, i.e. 5.4 minutes,
are indicated by vertical dashed line.
 }
 \label{dotf1fig8}
\end{figure}

\begin{figure}
\centering
\includegraphics [angle=0,scale=0.8]{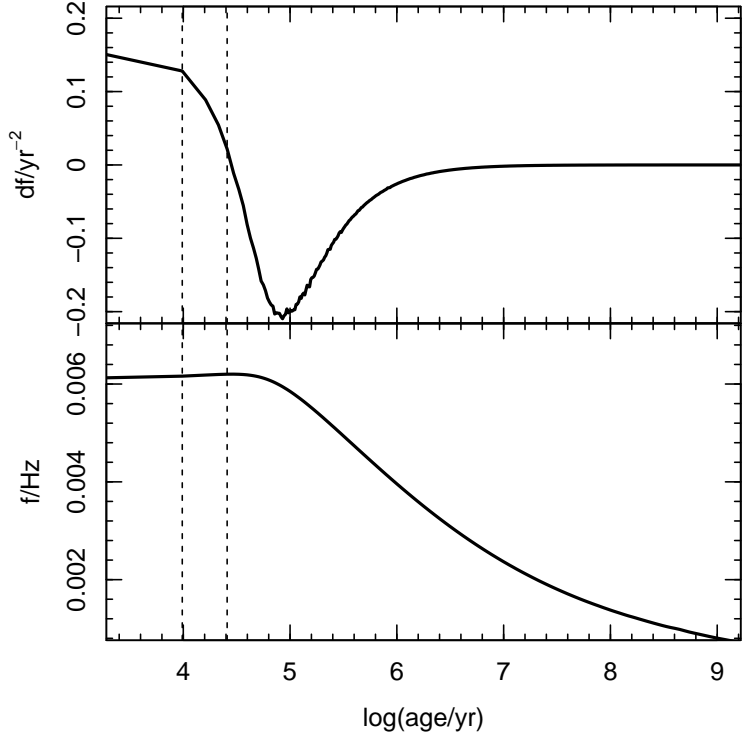}
\caption{Similar to Fig.~\ref{dotf1fig8}, the chirp and frequency evolution panel.
A possible region for HM\,Cnc is in between two vertical dashed lines.
 }
 \label{dotf2fig9}
\end{figure}

In a recent study of the implications of accreting double white dwarf binaries for LISA, \citet{Kremer2017} note that there are no observed DWDs with negative chirps. The best AM\,CVn candidate, HM\,Cnc, has a positive chirp suggesting that the GW radiation dominates the mass transfer. While \citet{Deloye2006} argue for a low turn-on timescale for mass transfer which would allow the gravitational radiation to dominate, \citet{Kremer2017} find that the turn-on timescale should be very short ($\sim 100\,\rm{yr}$) and hence that the 2$\times$HeWD model
is implausible for HM\,Cnc.

Fig.~\ref{dotf1fig8} shows the evolution of the  chirp (rate of frequency change) and mass-transfer rate with period for
the model of a $0.30$+$0.15\Msolar$ 2$\times$HeWD, with masses believed to be similar
to HM\,Cnc.
Combined with the observed orbital period of $5.4\,\rm{min}$, we
infer  $\dot{M} \approx 10^{-7}\,\Msolar\,\rm yr^{-1}$,
which is similar to that inferred in previous work, e.g.\citet{Bildsten2006}.
However, the chirp is about half the value $3.63\pm 0.06\times 10^{-16}\,{\rm Hz\,s}^{-1}$
reported by \citet{Strohmayer2005}.
Fig.~\ref{dotf2fig9} shows the chirp and frequency evolution with time.
The turn-on timescale is about $2.5\times10^4\,\rm{yr}$. A possible
region for HM\,Cnc is in between two vertical dashed lines, within which the
period (5.4 minutes) is constant  and the chirp is positive.
With such characteristics, LISA observations will be crucial in resolving the questions of the
formation channel.

\section{Conclusion}
We have computed the evolution of double  helium white dwarf  binaries covering a wide mass range  with mass ratios $q<2/3$, and including the detailed evolution of both stars.
Two different evolutionary pathways were considered, i.e. for low-mass ratio pairs to show stable Roche-lobe overflow (RLOF) over several Gyr, and for high-mass ratio pairs to form mergers.
The response of the accretor to its increasing mass has a modest effect on the boundary between RLOF pairs and mergers, but otherwise we have found negligible effect on the overall evolution of the system.

The value $q_{\rm crit}=2/3$ was calculated assuming a simple mass--radius relation for the HeWDs.
Generally speaking, such relations are derived for cold white dwarfs.
There are likely to be instances when the HeWD structure departs significantly from this condition.
Hence, some pairs of 2$\times$HeWDs with $q>2/3$ may evolve through stable Roche lobe overflow and not merge.
However, the most recent version of \texttt{MESA} does not provide a stable calculation for such models.
This will be examined carefully in future work.
Furthermore, we have not yet excluded He star+CO WD and He+CO WD binaries
as progenitors for these systems; this will require additional models.

From the models, an orbital period--donor-mass relation has been derived and used to estimate the component masses for eight AM\,CVn stars.
From radial velocity arguments, the models provide  evidence that central spikes in the triple-lined emission spectra of J1240, GP\,Com and V396\,Hya originate from or close to the accretor in 2$\times$HeWD systems.
In terms of surface composition, our 2$\times$HeWD models show surface $\rm{N/C}$ and $\rm{N/O}$ ratios very similar to GP\,Com, the only AM\,CVn in which such measurements have been made.

As AM\,CVn stars are amongst the most likely resolvable sources of Galactic gravitational wave radiation, we recalculate the predicted signal using our new mass estimates for the eight suspected 2$\times$HeWD AM\,CVn systems.
We confirm that the stars HM Cnc, AM CVn and V406 Hya are potentially detectable by LISA; the GW strain-amplitude would confirm the masses and hence possible origin for these systems.

Future work will examine the boundary between merger and stable RLOF channels in more detail, address its impact on the predicted number densities for AM\,CVn stars in the Solar neighbourhood, and compare with models for stable RLOF in He+CO WD binaries.

\normalem
\begin{acknowledgements}
We'd like to thank you the helpful suggestions and comments of referee to improve the manuscript.
This work is supported by the CAS ¡¯Light of West China¡¯ program (2015-XBQN-A-02),
the grants 10933002, 11703001 and 11273007 from the National Natural Science Foundation of China,
the Joint Research Fund in Astronomy (U1631236) under cooperative agreement between
the National Natural Science Foundation of China (NSFC) and Chinese Academy of Sciences (CAS),
the Strategic Priority Research Program of the Chinese Academy of
Sciences (No.XDB2304100), the China Postdoctoral Science Foundation,
and the Fundamental Research Funds for the Central Universities.
The Armagh Observatory and Planetarium is supported by a grant from the Northern Ireland Department for Communities.
CSJ and PDH acknowledge support from the UK Science and Technology Facilities Council (STFC) Grant No. ST/M000834/1

\end{acknowledgements}

\bibliographystyle{raa}
\bibliography{mybib}

\end{document}